\documentclass[apj]{emulateapj}
\usepackage[utf8]{inputenc}
\usepackage{lipsum}
\usepackage{amsmath, amssymb, bm}
\usepackage{balance}
\usepackage{multirow}
\usepackage{color}
\usepackage{enumitem}

\usepackage[usenames,dvipsnames]{xcolor}
\usepackage{hyperref}
\hypersetup{
    colorlinks = true,
    citecolor = {MidnightBlue},
    linkcolor = {BrickRed},
    urlcolor = {BrickRed}
}

\usepackage[caption=false]{subfig}
\usepackage{varwidth}

\begin{document}
\submitted{\today}
\title{The Effects of Bandpass Variations on Foreground Removal Forecasts for Future CMB Experiments}
\author{
J.~T.~Ward\altaffilmark{1},
D.~Alonso\altaffilmark{3},
J.~Errard\altaffilmark{2},
M.~J.~Devlin\altaffilmark{1},
M.~Hasselfield\altaffilmark{4},
}
\altaffiltext{1}{Department of Physics and Astronomy, University of Pennsylvania, Philadelphia, PA 19104, USA}
\altaffiltext{2}{AstroParticule et Cosmologie, Univ Paris Diderot, CNRS/IN2P3, CEA/Irfu, Obs de Paris, Sorbonne Paris Cit\'e, France}
\altaffiltext{3}{University of Oxford, Denys Wilkinson Building, Keble Road, Oxford OX1 3RH, UK}
\altaffiltext{4}{Pennsylvania State University, 520 Davey Lab, University Park, PA 16802, USA}

\begin{abstract}
Time-dependent and systematic variations in the band gain and central frequencies of instruments used to study the Cosmic Microwave Background are important factors in the data-to-map analysis pipeline.  If not properly characterized, they could limit the ability of next-generation experiments to remove astrophysical foreground contamination.  Uncertainties include the instrument detector band, which could systematically change across the focal plane, as well as the calibration of the instrument used to measure the bands.  A potentially major effect is time-dependent gain and band uncertainties caused by atmospheric fluctuations.  More specifically, changes in atmospheric conditions lead to frequency-dependent changes in the atmospheric transmission which, in turn, leads to variations in the effective gain and shifts in the central frequency of the instrument's bandpass.  Using atmospheric modeling software and measured ACTPol bandpasses, we simulate the expected variations in band gain and central frequency for 20, 40, 90, 150, and 240~GHz bands as a function of precipitable water vapor, observing angle, and ground temperature.  Combining these effects enables us to set maximum and minimum limits on the expected uncertainties in band gain and central frequency over the course of a full observing season.  We then introduce the uncertainties to parametric maximum-likelihood component separation methods on simulated CMB maps to forecast foreground removal performance and likelihoods on the tensor-to-scalar ratio r.  We conclude that to confidently measure a $\sigma(r=0) \sim 10^{-3}$ with a bias on the recovered $r$ under control, the limit on the uncertainty in the relative gain of the bandpass must be less than 2\% and the limit on the uncertainty in the central frequency of the bandpass must be less than 1\%.  The time variability of these parameters must be fully understood to realize the full impact of these uncertainties.  We also comment on the possibility of self-calibrating bandpass uncertainties.  
\end{abstract}

\section{Introduction}
Cosmic Microwave Background (CMB) studies are at the beginning of a new era, with Stage-3 experiments such as the Atacama Cosmology Telescope (ACT) \citep{2016SPIE.9910E..14D}, the Simons Array \citep{2016JLTP..184..805S}, CLASS \citep{2014SPIE.9153E..1IE}, SPT-3G \citep{2014SPIE.9153E..1PB}, BICEP3 \citep{2014SPIE.9153E..1NA}, and The Simons Observatory (SO) \footnote{\url{https://simonsobservatory.org/}} already in progress. Additionally, planned collaborative efforts such as CMB-S4 \citep{2016arXiv161002743A} are on the horizon.  These instruments will increase the total number of detectors on the sky by a factor of 10, driving our sensitivity to the faint CMB radiation to unprecedented levels.  One of the primary goals of these instruments is to measure the tensor-to-scalar ratio $r$, obtained through measurements of the B-mode polarization in the CMB at large angular scales.  

\begin{figure}[h] %  figure placement: here, top, bottom, or page
   \centering
   \includegraphics[width=3.25in]{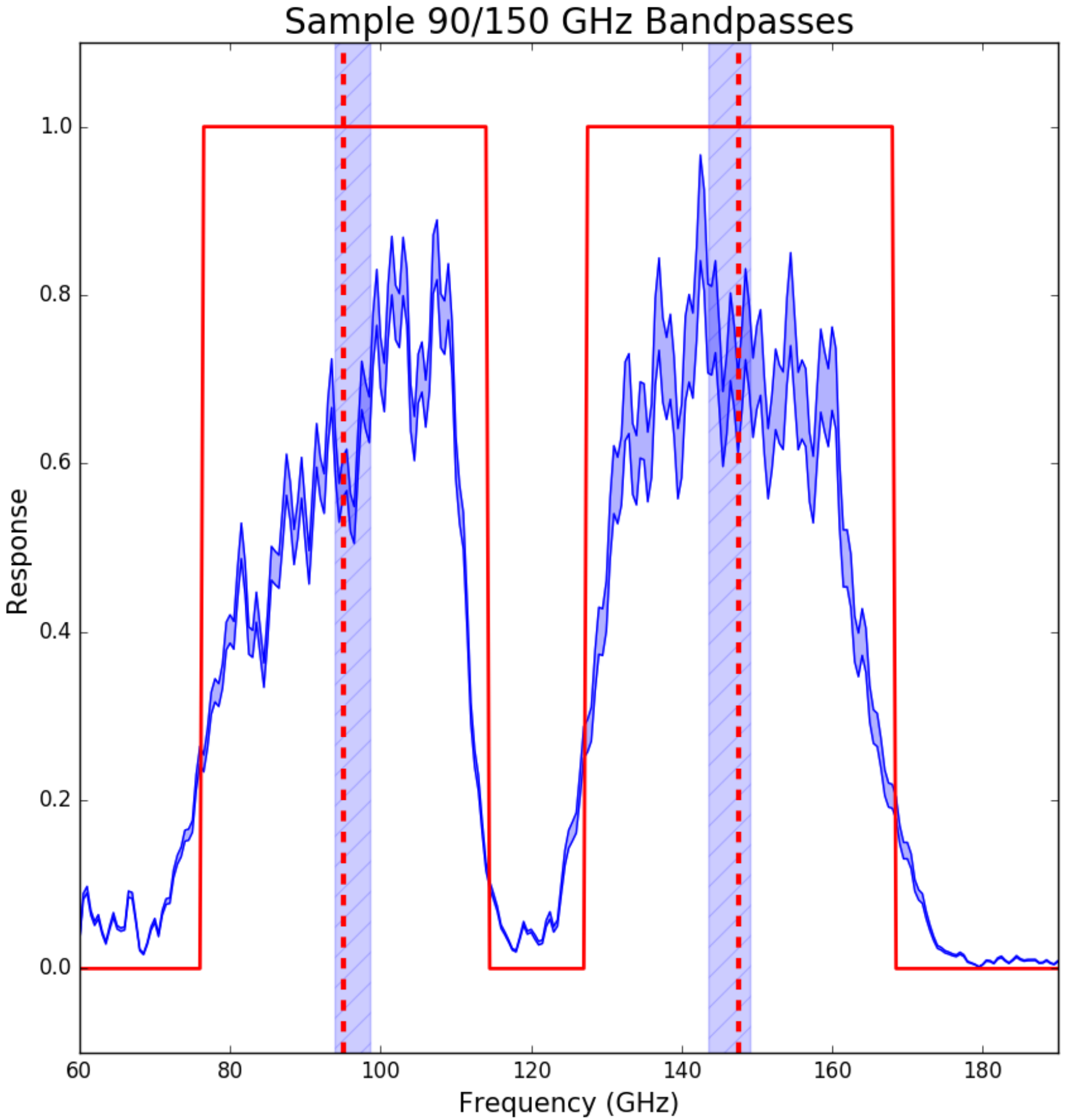} 
   \caption{Forecasting simulations typically utilize idealized bandpass features as shown in red.  This assumes that the bandpasses have top-hat profiles with well-known band centers.  However, the measured instrument bandpass typically exhibits what is shown in blue.  The profile is irregularly shaped and there is a level of uncertainty in both the band central frequency and gain (represented by the thickness of the shaded blue regions).}
   \label{Ideal_vs_real_bands}
\end{figure}

\begin{figure*}[t] %  figure placement: here, top, bottom, or page
   \centering
   \includegraphics[width=7.25in]{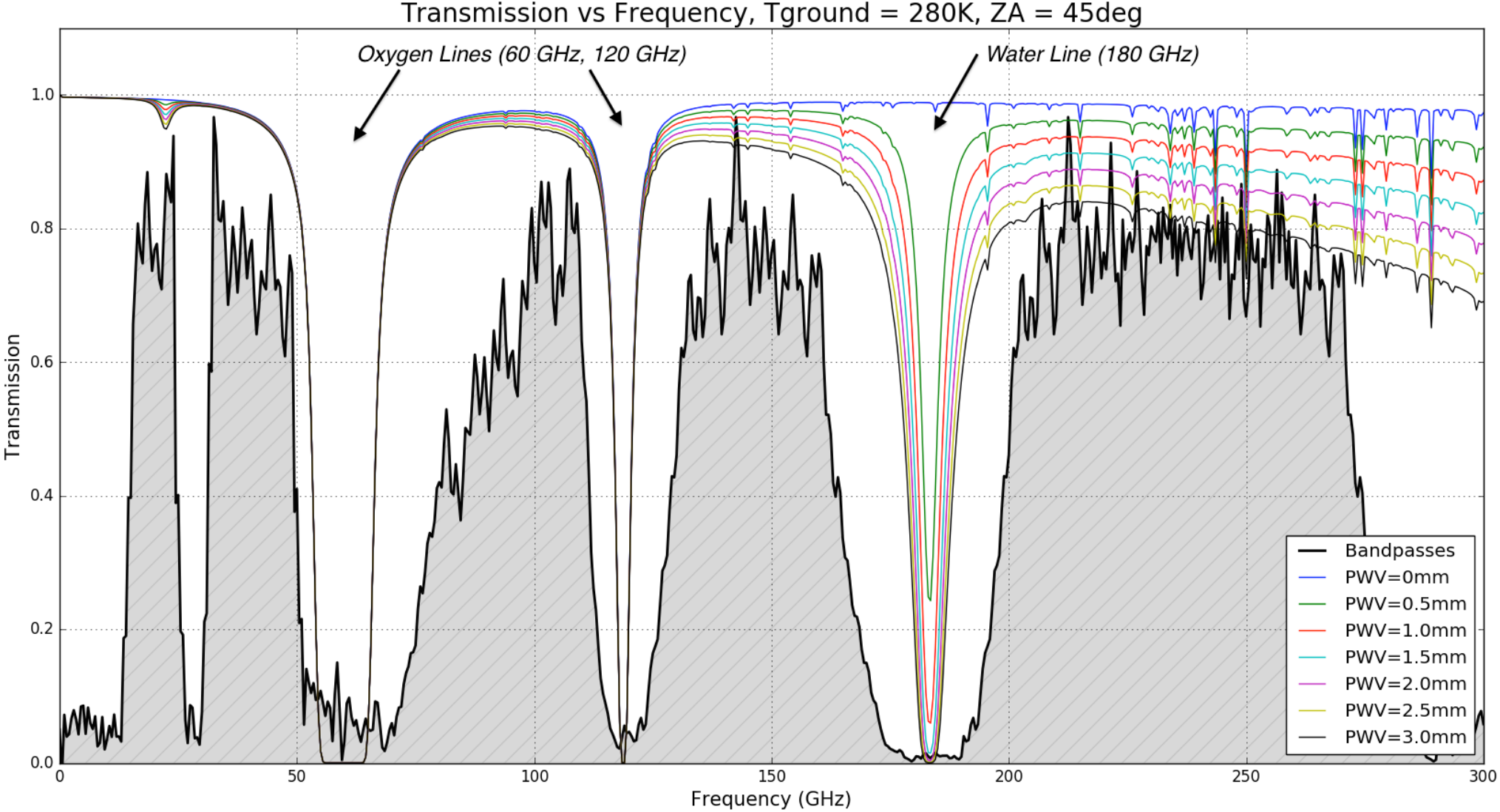} 
   \caption{Simulated instrument bandpasses for 20, 40, 90, 150, and 240~GHz bands.  The bands are modeled based on FTS measurements made on Advanced ACTPol detector arrays \citep{2016ApJS..227...21T} and are shown in black and shaded grey.  The colored lines represent atmosphere transmission lines for different values of precipitable water vapor (PWV), with ground temperature and zenith angle held at $280$~K and $45$ degrees, respectively.  The decrease in transmission as PWV increases suppresses the total gain in each band.  This suppression is also frequency-dependent, which leads to a shift in the central frequency of each band.  Notice how the O2 line impinges on the 40GHz band.  While this does have a significant offset in the band gain and central frequency, it does not vary much with changing PWV.}
   \label{Transmission_lines}
\end{figure*}

Although sensitivity forecasts indicate upcoming instruments will measure a $\sigma(r=0) \sim 10^{-3}$, the task of removing polarized foregrounds from CMB maps is a challenging and essential step in confirming a detection of $r$.  In order to remove the dust and synchrotron signals, they must be mapped over a range of frequencies at high sensitivity and measured accurately enough to suppress them by several orders of magnitude.  In addition, instrument systematics must be precisely characterized as any uncertainty can quickly affect the performance of cleaning methods such as the parametric maximum-likelihood methods (e.g. \cite{2008ApJ...676...10E,2009MNRAS.392..216S,2009ApJ...701.1804D,2012PhRvD..85h3006E,2016MNRAS.458.2032R,2017PhRvD..95d3504A}).  One of the key input parameters for foreground removal techniques is the information regarding the instrument's bandpass, specifically the absolute gain and central frequency \cite{2016A&A...594A..10P}.  Unfortunately, these values can be difficult to measure with the required precision and can vary over time depending on a wide range of factors.   

For the scope of this paper, we focus on bandpass gain and central frequency variations brought about by changing atmospheric conditions such as precipitable water vapor (PWV), observing angle, and ground temperature.  These variations yield an estimate of uncertainties in the band gain and central frequency.  Furthermore, we perform component separation techniques using gain and center values with the computed uncertainties rather than the idealized values shown in Figure \ref{Ideal_vs_real_bands}, which assumes top-hat band profiles with well-defined band centers.

\section{Modeling Band Variations}
Every CMB experiment will feature an inherent bandpass that is defined by the instrument's optical elements, feedhorns, and detectors.  The properties of the bands can be measured both in the laboratory and in the field with a Fourier Transform Spectrometer (FTS).  While the precision of these measurements has been sufficient to date, they may not meet the sensitivity levels required for the planned instruments.  In addition, the gain and central frequency of each band can vary during observations due to changing atmospheric conditions.  We estimate the amplitude of atmosphere-induced variations by simulating atmospheric transmission lines and instrument bandpasses and calculating the resultant changes in gain and central frequency for each band.

\subsection{Bandpasses and Atmosphere Transmission Lines}

\begin{figure*}[t] %  figure placement: here, top, bottom, or page
   \centering
   \includegraphics[width=\linewidth]{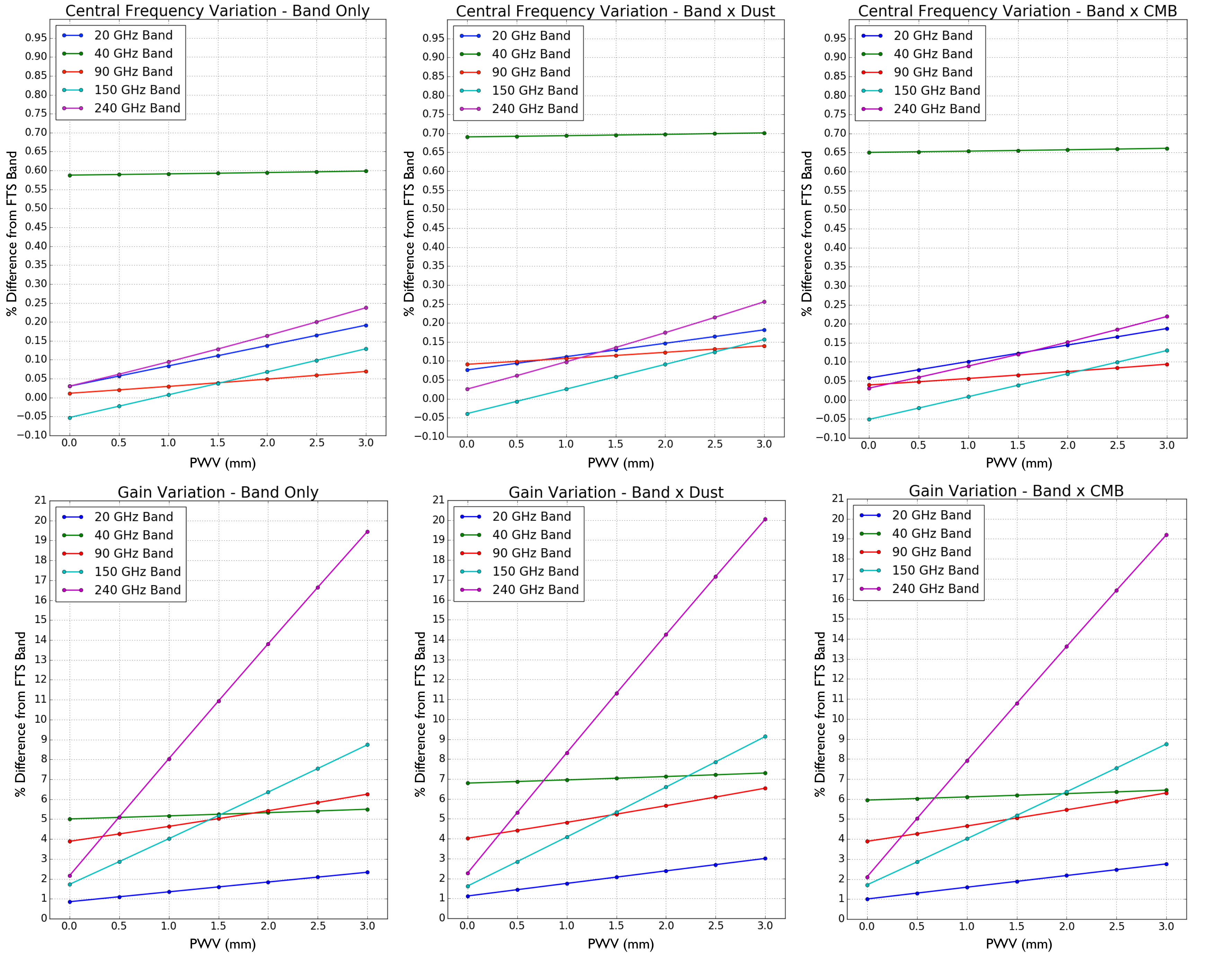} 
   \caption{Percent change in gain variation and central frequency as a function of PWV for the instrument band only, the band multiplied by a dust spectrum, and the band multiplied by a CMB spectrum.  Going from a PWV value of 0 to 3~mm, a maximum change in gain of about 20 percent is evident in the 240~GHz band and a maximum change in the central frequency of about 0.7 percent is evident in the 40~GHz band.  We set the upper limit of the PWV to 3mm because any data taken with a PWV greater than this value is not used.}
   \label{PWV_changes}
\end{figure*}

The first step towards modeling bandpass variations is to start with realistic instrument bands.  We achieve this by referencing actual measurements taken of ACTPol detectors.  The 20, 40, and 240 GHz bands used for the model are FTS measurements taken on Advanced ACT single pixels at the National Institute of Standards and Technology (NIST).  The 90 and 150 GHz bands are FTS measurements taken of full ACTPol detector arrays that are currently deployed on the Atacama Cosmology Telescope in Chile.  All five of the bands are plotted together in Figure \ref{Transmission_lines} and are outlined in black and shaded grey.  

To generate atmosphere transmission spectra, we use the Atmospheric Model {\tt am} \citep{paine_scott_2017_438726}.  A specific template is used to simulate the conditions on the Chajnantor Plateau in Northern Chile, a location that is home to both existing and future CMB experiments. {\tt am} produces opacity data based on user inputs such as PWV, zenith angle, ground temperature, and frequency range.  The opacity is then converted to transmission by using the relation
\begin{equation}\nonumber
T(\nu) = e^{-\tau(\nu)}, 
\end{equation}
where T is transmission and $\tau$ is opacity.  Figure \ref{Transmission_lines} shows transmission spectra plotted for PWV values from 0 to 3 millimeters in increments of 0.5mm.  The atmosphere-included bands are calculated by multiplying the instrument bands by the simulated transmission spectra, resulting in a final band for each PWV, zenith angle, and ground temperature value.  

\subsection{Computing Gain and Central Frequency Variations}

\begin{figure*}[t] %  figure placement: here, top, bottom, or page
   \centering
   \includegraphics[width=\linewidth]{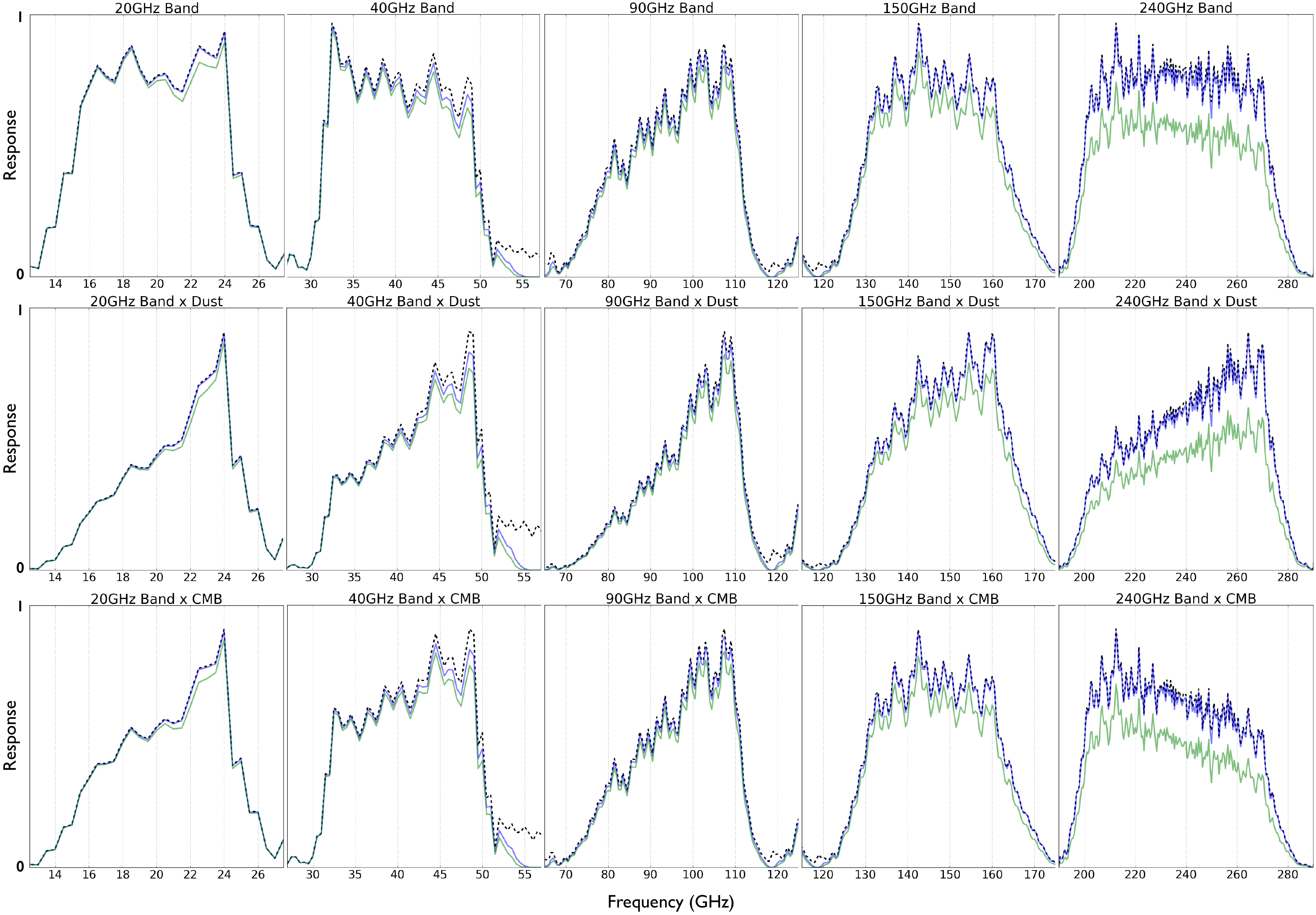} 
   \caption{Minimum (PWV=0~mm, ZA=$30^\circ$, $T_{ground}$=290~K) and maximum (PWV=3~mm, ZA=$60^\circ$, $T_{ground}$=250~K) effects of the atmosphere for each of the five simulated frequency bands.  The top row shows the instrument band, the middle row shows band x dust, and the bottom row shows band x CMB.  The blue represents the minimum effect, the green represents the maximum effect, and the black dashed line shows the original band as measured by the FTS.  The y-axis of the plots goes from zero to one and represents the normalized transmission of the band, while the x-axis is the frequency in gigahertz.}
   \label{minmax_bands}
\end{figure*}

The gain for each frequency band is calculated by numerically integrating the area under the curve for a given band width.  The band edges are set to be where the band response hits the noise floor, but in practice the effective bands may be narrower.  

\begin{table}[]
\centering
\label{my-label}
\begin{tabular}{|l|r|r|r|}
\hline
\multicolumn{4}{|c|}{\textbf{Band Properties}}                                                                      \\ \hline
Frequency Band & \multicolumn{1}{l|}{Min (GHz)} & \multicolumn{1}{l|}{Max (GHz)} & \multicolumn{1}{l|}{Width (GHz)} \\ \hline
20 GHz         & 10                             & 30                             & 20                               \\ \hline
40GHz          & 28                             & 55                             & 27                               \\ \hline
90 GHz         & 65                             & 125                            & 60                               \\ \hline
150 GHz        & 120                            & 180                            & 60                               \\ \hline
240 GHz        & 190                            & 290                            & 100                              \\ \hline
\end{tabular}
\caption{Integration limits for the simulated bands used in the analysis.  The values were chosen based on the response cutoff for each band, ensuring that the maximum bandwidth is utilized.  However, the effective bandwidths are narrower.}
\label{band_properties}
\end{table}

For this study, it is important to capture the behavior across the entire possible bandwidth for each band in order to place upper limits on the total atmospheric effects.  The chosen integration limits are outlined in Table \ref{band_properties}. 

The central frequency of each band is computed by taking a weighted average across the band width, where the weights are the transmission of the band in each frequency bin.  This method accounts for the specific shape of the band in addition to using only the band limits to calculate the effective center.  Once the gain and central frequency are computed for each transmission line, the atmosphere-included bands are compared to the instrument-only bands to quantify the changes as a percent difference from the original FTS measured band. 

\subsection{PWV Results}
Of the three atmospheric effects considered in the model, PWV is the strongest driver of variations in the properties of instrument bandpasses.  The results for percent changes in gain and central frequency of the band for incremental steps of PWV are shown in Figure \ref{PWV_changes}.  We use a range of 0 to 3~mm because data is typically not used when the PWV exceeds 3~mm, and radiometer data indicates that the PWV can vary across this entire range throughout an observing season.  We obtain a linear relationship between PWV and the percent change in gain and central frequency.  The variations in the gain are approximately an order of magnitude larger than the variations in the band center, however the central frequency shifts are still significant.  For the gain variation across a full observing season, we estimate a range of 1.0 to 3.0 percent change for 20~GHz, 6.8 to 7.4 percent change for 40~GHz, 4.0 to 6.6 percent change for 90~GHz, 1.6 to 9.2 percent change for 150~GHz and 2.3 to 20 percent change for 240~GHz.  For the central frequency variation across a full observing season, we estimate a range of 0.05 to 0.19 percent change for 20~GHz, 0.65 to 0.66 percent change for 40~GHz, 0.04 to 0.1 percent change for 90~GHz, -0.05 to 0.13 percent change for 150~GHz and 0.03 to 0.22 percent change for 240~GHz.  See Table \ref{tab:minmax_center} for the full set of variations.

\subsection{Total Atmosphere Results}
The effects on instrument bandpasses from varying the observing angle and ground temperature at fixed values of PWV are significantly smaller than the changes brought on by varying PWV values.  Therefore, the individual results for the variation of observing angle and ground temperature are not presented as was done for PWV in Figure \ref{PWV_changes}.  However, we want to set upper and lower limits on the total effects seen from the atmosphere by comparing the best and worst case combinations of PWV, zenith angle, and ground temperature parameters.  After researching the typical observing conditions on the Chajnantor Plateau in Chile (the location of CMB experiments such as ACT, CLASS, POLARBEAR, and the Simons Array), we find that over a typical observing season the PWV ranges from 0 to 3 mm, the observing angle ranges from 30 to 60 degrees, and the ground temperature ranges from 250 to 290 Kelvin.  The optimal conditions we expect to see during an observing season are PWV=0, ZA=30$^\circ$, and T$_{\rm ground}$=290K.  The worst conditions that we would see during an observing season while still using the data for the final CMB maps are PWV=3, ZA=60$^\circ$, and T$_{\rm ground}$=250K.  These values were used to generate best and worst-case scenario transmission lines to solve for band variations as described in the previous sections.  

\vspace{1mm}
\begin{table}
 \centering
 \resizebox{\linewidth}{!}{%
 \begin{tabular}{|c|l l|l l|l l|}
  \hline
  \multicolumn{7}{|c|}{Min and Max Percent Changes - Band Gain} \\
  \hline
  \multirow{3}{*}{Band} 
      & \multicolumn{2}{c|}{Band}
          & \multicolumn{2}{c|}{Dust}
              &\multicolumn{2}{c|}{CMB}\\             \cline{1-7}
  & Min & Max & Min & Max & Min & Max \\  \hline
   20GHz & 0.73\% & 3.22\% & 1.00\% & 4.08\% & 0.88\% & 3.77\% \\      
   40GHz & 4.24\% & 7.47\% & 5.83\% & 9.71\% & 5.07\% & 8.66\% \\      
   90GHz & 3.19\% & 8.98\% & 3.32\% & 9.35\% & 3.20\% & 9.04\% \\      
   150GHz & 1.41\% & 12.8\% & 1.33\% & 13.3\% & 1.40\% & 12.8\% \\      
   240GHz & 1.77\% & 27.8\% & 1.86\% & 28.6\% & 1.73\% & 27.45\% \\      \hline
 \end{tabular}}
\label{tab:minmax_gain}
\end{table}
\begin{table}
 \centering
 \resizebox{\linewidth}{!}{%
 \begin{tabular}{|c|l l|l l|l l|}
  \hline
  \multicolumn{7}{|c|}{Min and Max Percent Changes - Band Center} \\
  \hline
  \multirow{3}{*}{Band} 
      & \multicolumn{2}{c|}{Band}
          & \multicolumn{2}{c|}{Dust}
              &\multicolumn{2}{c|}{CMB}\\             \cline{1-7}
  & Min & Max & Min & Max & Min & Max \\  \hline
   20GHz & 0.03\% & 0.26\% & 0.07\% & 0.23\% & 0.05\% & 0.25\% \\      
   40GHz & 0.51\% & 0.78\% & 0.61\% & 0.89\% & 0.57\% & 0.85\% \\      
   90GHz & 0.01\% & 0.10\% & 0.08\% & 0.19\% & 0.03\% & 0.13\% \\      
   150GHz & -0.04\% & 0.20\% & -0.03\% & 0.24\% & -0.04\% & 0.20\% \\      
   240GHz & 0.02\% & 0.36\% & 0.02\% & 0.39\% & 0.02\% & 0.33\% \\      \hline
 \end{tabular}}
\caption{Minimum and maximum percent changes in band gain (top) and central frequency (bottom).}
\label{tab:minmax_center}
\end{table}

The results for each band and each observed source are shown in Figure \ref{minmax_bands}.  Note that the amount of change is different depending on the band, leading to uneven shifts across all frequencies.  Also note that the band profile is different depending on the observed source, which leads to differences in how the atmosphere affects the gain and central frequency of the pure band or for observations of a CMB-like or a dust-like source. The minimum and maximum expected percent changes in band gain and central frequency across all bands for each observed source are listed in Table \ref{tab:minmax_center}. 

\section{Propagation of uncertainties to tensor-to-scalar measurements}
Here we describe the methods used to propagate the effects of variations in the gains and central frequencies of all bands into the final uncertainties in the measurement of the tensor-to-scalar ratio $r$ after foreground cleaning.\\

\subsection{Methodology}
\label{ssec:methodology_xForecast}

\begin{figure*}[t] %  figure placement: here, top, bottom, or page
   \centering
   \includegraphics[width=5in]{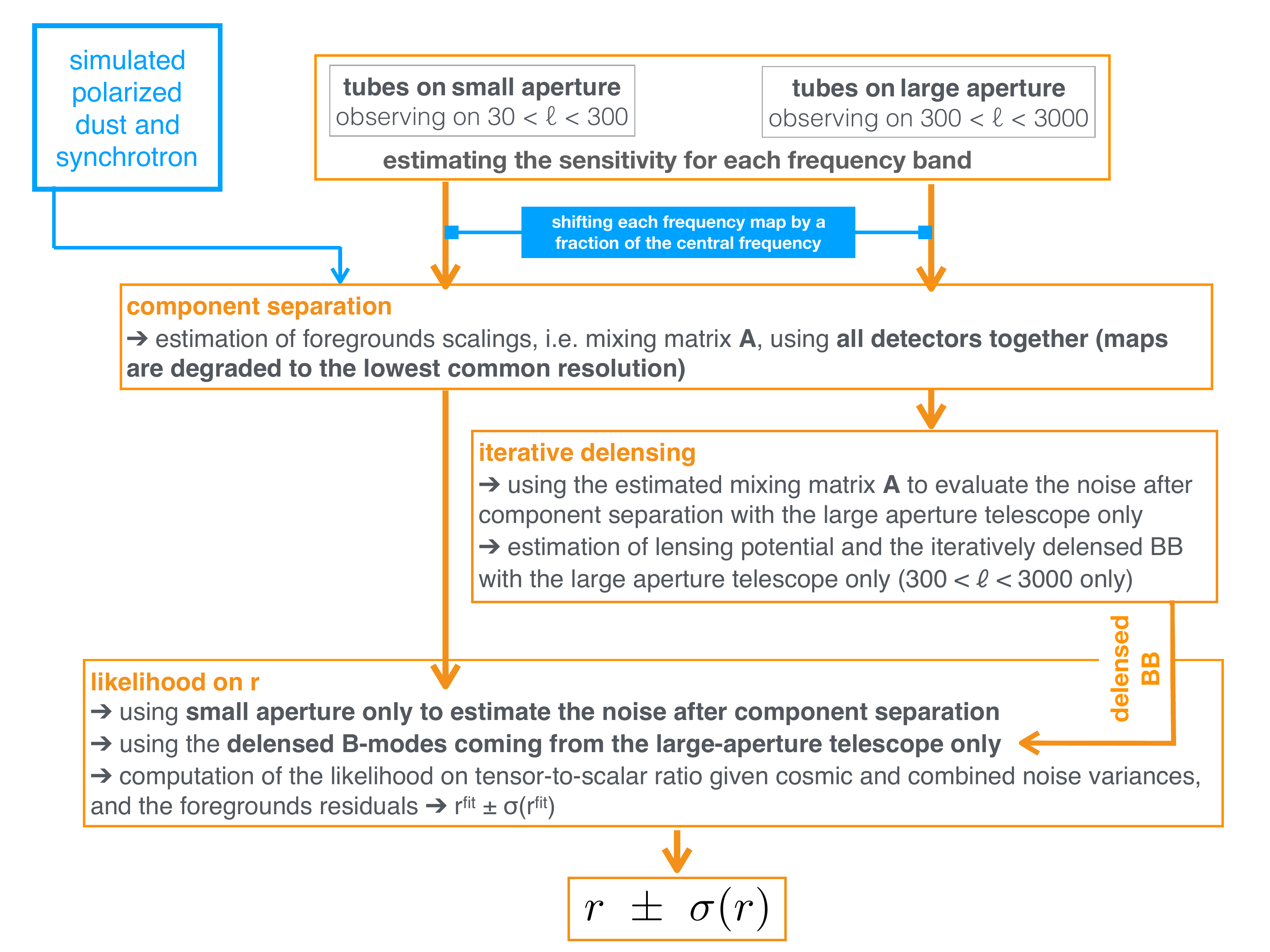} 
   \caption{Schematic of the xForecast wrapper to study bandpasses effects. See section~\ref{ssec:methodology_xForecast} for details.}
   \label{fig:schematic_xForecast}
\end{figure*}

We exploit the formalism described in \cite{2016PhRvD..94h3526S}, which proposes a framework to forecast both bias and uncertainty on the tensor-to-scalar ratio after cleaning synchrotron and dust. This framework assumes the map-level parametric component separation described below, marginalizing over foreground spectral parameters and amplitudes. We start from a set of sky simulations generated using PySM (described in \cite{2017MNRAS.469.2821T}), containing a single dust component with a modified black-body spectrum and polarized synchrotron emission with a power-law frequency dependence. The amplitudes and spectral parameters for both foreground sources are based on the latest Planck component-separated COMMANDER maps \citep{2016A&A...594A..10P}. In addition to the foregrounds, we simulate the CMB B-mode emission in the absence of primordial B-modes, and thus any recovered $r\neq0$ after component separation would denote a bias. 

We consider a typical CMB-S3/4 observatory, with hybrid telescopes i.e. a combination of small and large aperture telescopes having the same set of five frequencies (20, 40, 90, 150, 240) GHz with the following sensitivities in polarization: (10.9, 7.5, 1.7, 2.1, 5.1) $\mu$K-arcmin. We assume that data from the large aperture telescope (6-m mirror size providing a $\sim$1.5' FWHM at 150~GHz) is used to estimate the lensing potential over the observed field, which provides an estimate of the lensing B-mode maps given an observation of the E-modes. This information is used in the likelihood on tensor-to-scalar ratio in order to reduce the variance induced by lensing B-modes.
In this work, we focus on the cleanest $f_{\rm sky}=5\%$ of the sky observable from the Southern Hemisphere in terms of foregrounds, as determined in \cite{2017PhRvD..95d3504A}.

Given a set of frequency maps, we then exploit a foreground cleaning algorithm based on parametric component separation, as detailed in \cite{2016PhRvD..94h3526S}. In this study, we follow the reasoning described in Figure~\ref{fig:schematic_xForecast}. 
We assume that two telescopes, having a small and a large apertures, and having the same sensitivities as quoted above, are first combined to estimate the spectral parameters, specifically $\beta_s$ for synchrotron and $\beta_d$ for dust. This estimation is done via the optimization of a so-called spectral likelihood, analytically averaged over CMB and noise realizations, assuming constant spectral indices over the observed sky patch -- the foreground residuals induced by the spatial variability of the spectral indices turn out to be below the statistical uncertainty of the B-mode measurement.
 
Once the spectral parameters $\{\beta_d,\beta_s\}$ are estimated, we can compute the amplitude of both statistical and systematic foreground residuals, as detailed in \cite{2016PhRvD..94h3526S}. This is then used in a cosmological likelihood on $r$, which allows us to collect the CMB and noise-averaged $r\,\pm\,\sigma(r)$.
During the estimation of the likelihood on $r$, we look at two independent cases, without delensing (full lensed B-modes is used in the equations) and with iterative delensing as performed by the high-resolution, large aperture telescope.

Although the main results presented here were derived using the forecasting method described above, the consistency and robustness of these results were verified by comparison with an alternative power-spectrum-based component separation algorithm, similar to that implemented in \citep{2015PhRvL.114j1301B}. In this case, foregrounds and CMB are separated at the power-spectrum level by modelling the multi-frequency power spectra of all sky components. The method marginalizes over a set of 11 foreground parameters, including the amplitudes, spectral indices and power spectrum tilts of all components and an overall cross-correlation coefficient between synchrotron and dust. Although this method is less sophisticated than the map-based cleaning introduced above, it allows for a faster evaluation of the multi-frequency likelihood, and therefore we are able to Monte-Carlo over all foreground and CMB parameters with no approximations.

Both cleaning methods consist of a sky model (describing the physical parameters of the different components) and an instrument model, describing the bandpasses, beams and noise levels of the different frequency channels. In order to study the effect of bandpass systematics, we perturb the instrument model before component separation by introducing a Gaussian random shift in the central frequency and overall gain of each frequency channel $i$:
\begin{align}
%\centering
    	\nu_i \rightarrow \nu_i \times \left( 1 +  \mathcal{N}\left( 0,\sigma_\nu \right) \right),\\
    	  g_i \rightarrow g_i \times \left( 1 +  \mathcal{N}\left( 0,\sigma_g \right) \right),
\end{align}
where $\mathcal{N}(\mu,\sigma)$ is a Gaussian random number (independent for each frequency channel) with mean $\mu$ and standard deviation $\sigma$. The fiducial gains are $g_i=1$ and $\sigma_{\nu/g}$ parametrizes the uncertainty on either central frequencies or gains. Here we explore values $0<\sigma_{\nu/g}<5\%$.

Finally, we run 100 simulations for each value of $\sigma_\nu$ and $\sigma_g$, leading to $100$ estimates of CMB- and noise-averaged $r\,\pm\,\sigma(r)$.

\subsection{Results}\label{sec:results}

\begin{figure*}[t] %  figure placement: here, top, bottom, or page
   \centering
   \includegraphics[width=7.25in]{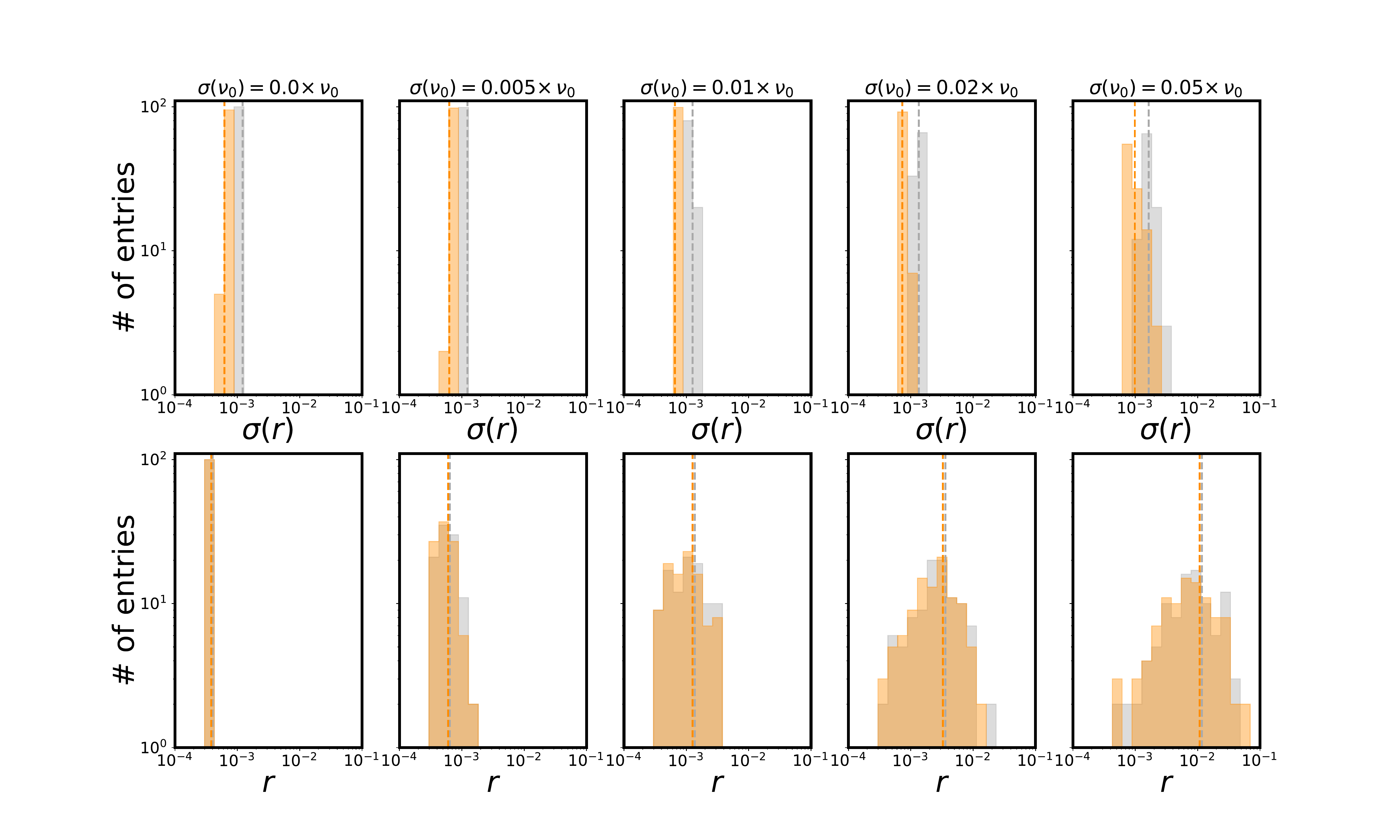} 
   \caption{Distribution of $\sigma(r)$ and $r$ as derived by xForecast, run for 100 simulated central frequencies uncertainties (see title of each column). The means of each distributions are shown as dashed colored vertical lines.}
   \label{fig:result_from_xForecast}
\end{figure*}

\begin{figure*}[t] %  figure placement: here, top, bottom, or page
   \centering
   \includegraphics[width=7.25in]{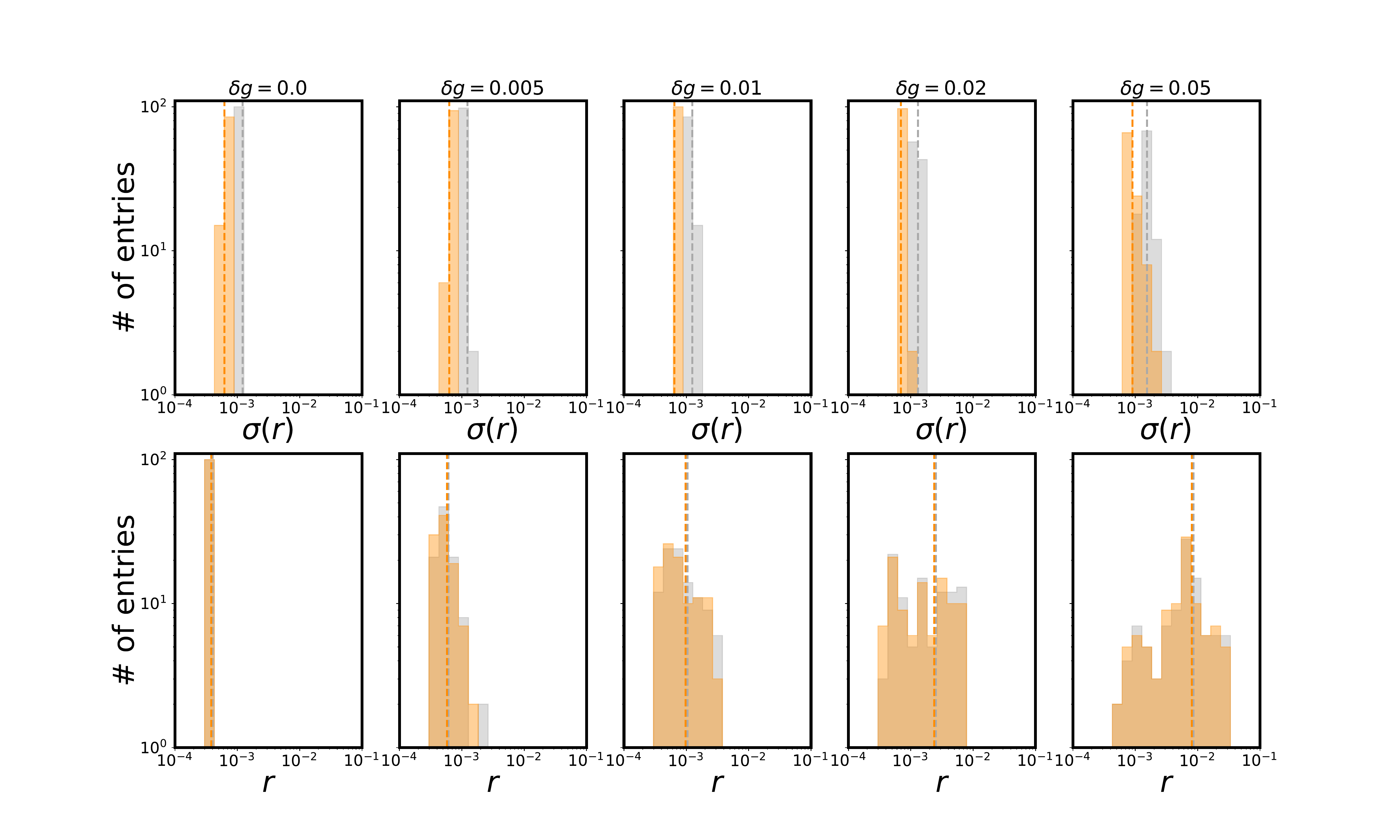} 
   \caption{Same as Fig.~\ref{fig:result_from_xForecast} but for relative gain uncertainties.}
   \label{fig:result_from_xForecast_gains_uncertainties}
\end{figure*}

The results for relative bandpass and gain uncertainties are shown in Figures \ref{fig:result_from_xForecast} and \ref{fig:result_from_xForecast_gains_uncertainties} respectively, following the analysis described above. In each figure, the first row shows the distribution of $\sigma(r)$, i.e. the inferred uncertainty on $r$ measurement, with (orange) and without (gray) including the iterative delensing coming from the large aperture telescope. $\sigma(r)$ does not depend strongly on the included systematic effects, as it is mainly driven by the noise after component separation, as well as from the variance from lensing B-modes. This is true until the foregrounds residuals exceed the total, cosmological B-modes signal: in this case, the variance from the foregrounds residuals contributes significantly to the overall error budget while estimating the tensor-to-scalar ratio. 

The second row of each figure shows the distribution of the bias on the tensor-to-scalar ratio $r$, as the input sky simulations have been assuming $r=0$. The fact that $r \sim 5\times 10^{-4}$ in the case of $\sigma(\nu_0) = 0.0 \times \nu_0$ or $\delta g = 0.0$ is due to the fact that our methods assume constant spectral indices, $\beta_d$ and $\beta_s$, whereas input simulated skies use spatially-varying spectral indices. However, since $r \ll \sigma(r)$ in both cases, even when including delensing, we do not consider extra degrees of freedom in the fit of the foreground frequency spectra, since doing so would significantly increase the noise in the final CMB map.

It is interesting to look at the significance of the bias on $r$, induced by foregrounds residuals, as a function of systematics amplitude. These results can be translated into bandpass calibration requirements by requiring that the uncertainty on these two systematics should only cause a bias on the tensor-to-scalar ratio $r \leq \sigma(r)$. Our results indicate that bandpass systematics must be kept within the following limits:
\begin{equation}\label{eq:req}
\Delta\nu_0/\nu_0 \lesssim 0.01,\hspace{12pt}\Delta g \lesssim 0.02.
\end{equation}
As described above, these results were verified by both of our foreground removal methods.

\begin{figure}
   \centering
   \includegraphics[width=0.47\textwidth]{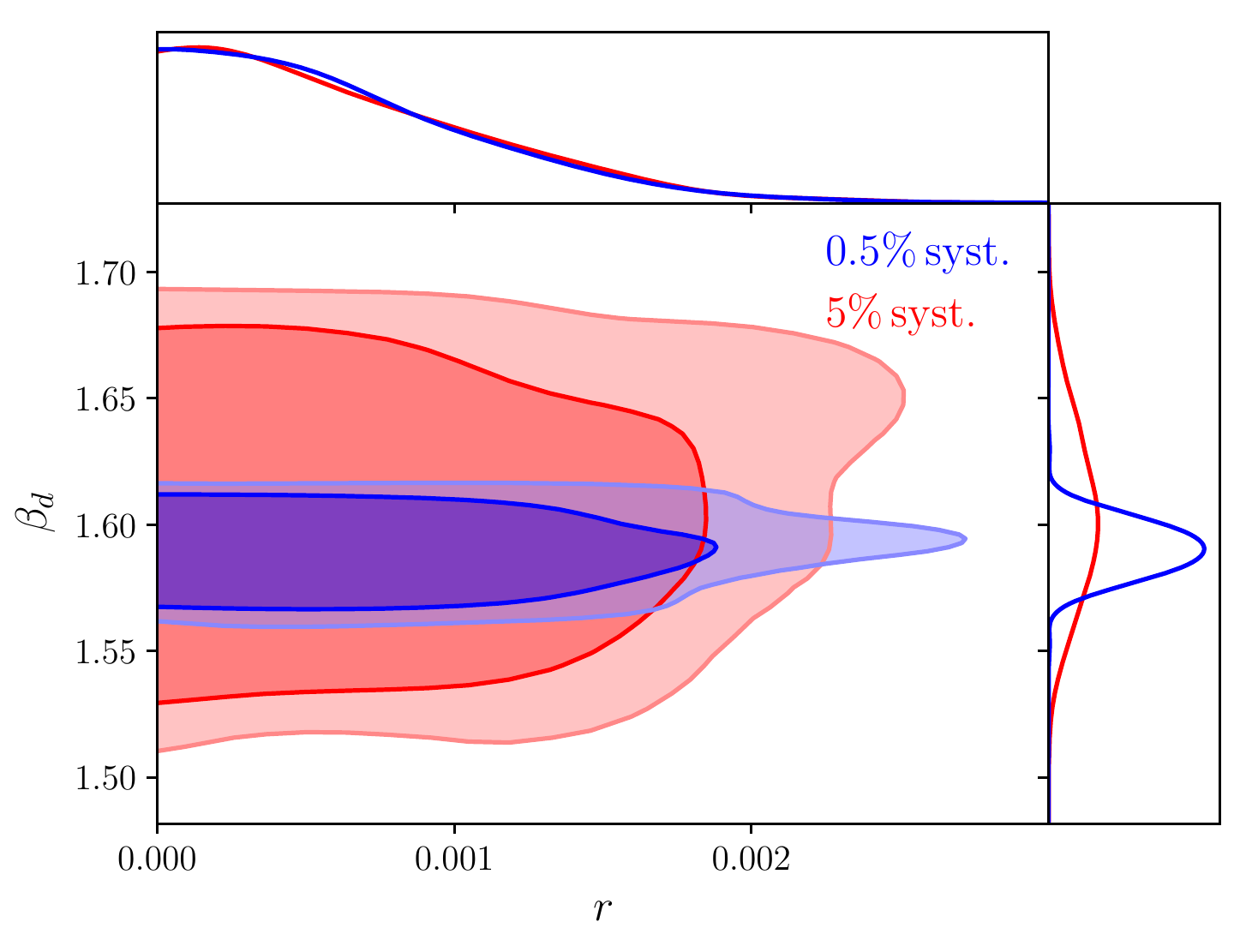} 
   \caption{2D contours showing the posterior uncertainties on the tensor-to-scalar ratio $r$ and on the dust spectral index $\beta_d$. 1$\sigma$ and 2$\sigma$ contours are shown two values ($5\%$ in red, $0.5\%$ in blue) of the Gaussian prior on the relative central frequencies of all 6 bands, which are also marginalized over. Although bandpass uncertainties significantly widen the uncertainties on foreground parameters, the final constraints on $r$ remain mostly unaffected.}
   \label{fig:lk_broaden}
\end{figure}

\subsection{Bandpass self-calibration}
In the previous section we have quantified the bias on $r$ associated to the shift in the best-fit sky model caused by a wrong instrument model in the presence of bandpass systematics. Our results show that percent-level accuracy in the calibration of global bandpass parameters will be needed to mitigate this bias beyond the 1$\sigma$ level. However, in the case of extreme band shifts this biased best-fit is likely to have a poor associated $\chi^2$. In the presence of residual bandpass systematics, the correct procedure would then be to marginalize over these systematics (frequency and gain shifts in our case) with a calibration-based prior. This procedure then produces an unbiased estimate of $r$ (assuming the correct sky model is used), at the cost of a potential increase in the final parameter uncertainties associated with the additional degrees of freedom. 

To explore this possibility, we have used our power-spectrum-based cleaning pipeline, including the frequency and gain shifts in all channels as additional free parameters with different priors. After marginalizing over the 10 bandpass parameters we observe that, on the one hand, the uncertainty on foreground parameters, such as the amplitudes and spectral indices of synchrotron and dust, do suffer significantly from the additional uncertainty on the bandpass systematics. On the other hand, however, the final uncertainties on the tensor-to-scalar ratio are more robust against these systematics, and can afford uncertainties of up to $\sim10\%$ on the bandpass parameters before $\sigma(r)$ starts degrading significantly with respect to the case of perfect bandpass calibration. This is explicitly shown in Figure \ref{fig:lk_broaden}, which displays the broadening of the final uncertainties in the $r$-$\beta_d$ plane assuming $5\%$ (red) and $0.5\%$ uncertainties on the relative central frequencies of all bands. These results are not completely unexpected: the CMB spectrum is known with a very good precision, unlike the case of foregrounds, and the fact that the CMB parameters are more robust to bandpass uncertainties is to be expected. In this respect, it is also worth noting that, although the parametric component separation methods used in this paper are more optimal than blind approaches such as ILC given the correct foreground model, they are also more sensitive to bandpass systematics, to which methods that only assume knowledge of the CMB spectrum are naturally more robust.

Nevertheless, it is important to note that our ability to self-calibrate the residual bandpass uncertainties as described above depends crucially on using a sufficiently accurate foreground model. To test this quantitatively, we have generated simulated multi-frequency power spectra using a slightly more complex foreground model, containing a synchrotron curvature coefficient of $C=0.2$, varying the dust temperature to $\Theta_d=30\,{\rm K}$ and introducing frequency decorrelation ($C_\ell(\nu_1,\nu_2)\propto\exp(-\xi\log^2(\nu_1/\nu_2))$, \citep{2017A&A...603A..62V}) in both components, with a correlation length $\xi=10$. Running the power-spectrum-based pipeline on these simulated data without accounting for any of these additional foreground degrees of freedom, we obtain biases in both the bandpass parameters (assuming a $5\%$ calibration error) and in the tensor-to-scalar ratio that oscillate between 2 and 5$\sigma$ significance. Therefore, given current foreground uncertainties, the recommended calibration requirements are those stated in Eq. \ref{eq:req}. These findings are, however, reassuring regarding the ability of future ground-based experiments to put strong constraints on the amplitude of primordial gravitational waves in the presence of residual bandpass uncertainties.

\section{Discussion}
In this paper we have quantified the impact of atmospheric effects on the bandpass of ground-based CMB experiments. The most relevant effect is the atmospheric PWV, which affects the band transmission far more than other observing conditions, such as zenith angle or ground temperature. We have shown that band gains can be substantially modified by atmospheric effects (see top of Table \ref{tab:minmax_center}), but it is important to note that other factors can have an effect on the gain and central frequency of the instrument bandpass.

To account for the gain attenuation introduced by the atmosphere, techniques are already implemented in the analysis pipelines of current CMB experiments.  For example, ACT uses planet observations to obtain absolute gain calibrations as a function of PWV and observing angle.  However, these calibrations exhibit scatter in the measured temperature of Uranus that is not well understood.  In addition to the atmosphere, detector parameters such as loading, bias, and bath temperature can all change how a detector will respond to the sky signal.  For ACT, the changes in calibration due to these effects are captured by taking bias steps (a direct measurement of the digital response of a detector to injection of power) to obtain detector conversions from DAC units to picoWatts.  The final calibrated ACT maps are compared to Planck results to determine the total gain uncertainty.  In the end, on-site calibration (e.g. using bias steps and planet observations) reduces the uncertainty in the final maps to the $\sim1\%$ level.  When the parameters affecting gain calibrations are properly accounted for, reaching the required gain uncertainty levels defined in this paper appears to be attainable.

Central frequencies are less affected by atmospheric observing conditions (see bottom of Table \ref{tab:minmax_center}), but as with gain, the atmosphere is not the only cause of band center shifts.  Detector environment parameters such as non-uniform on-chip filter fabrication, feedhorn manufacturing, and instrument systematics can all lead to band shifts on the per-detector level, creating a non-uniform focal plane.  Because the focal plane consists of thousands of detectors, an array-averaged band center is used in the analysis.  For ACT, and many other experiments, individual detector bands are measured at varying locations on the focal plane using a Fourier Transform Spectrometer (FTS) to obtain an average band center value.  The estimated uncertainty for ACT in the averaged band alone (not including atmosphere) due to measurement error in the FTS and the parameters listed above is approximately 2\% \citep{2016ApJS..227...21T}.  When combined with errors introduced by atmospheric effects, the total uncertainty exceeds the limit set by the analysis in Section 3.  This result suggests that it will be necessary to implement an atmospheric correction to the band central frequency.  If this correction is made assuming a PWV of 1.5~mm, the atmosphere-induced uncertainty in the band central frequency will be suppressed to approximately 0.1 percent.  At that level, the uncertainty due to atmospheric fluctuations will be sub-dominant to the FTS measurement uncertainty.  The correction can be calculated using the simulations outlined in Section 2.  Furthermore, the band center requirements can be fully satisfied by improving the band center measurements through better self-calibration of the FTS and characterizing a larger percentage, if not all, of the individual detectors in the focal plane.  Overall, reaching the required uncertainty limits in central frequency is achievable when pairing an atmospheric correction with improved bandpass measurements.

To determine these requirements, we have carried out simulated measurements of the tensor-to-scalar ratio $r$ on synthetic sky maps containing realistic foregrounds, CMB and bandpass systematics. If not accounted for, bandpass variations will induce a bias on the measured value of $r$, due to an imperfect cleaning of foregrounds. We have defined our requirement on bandpass calibration such that the associated bias on $r$ remain below the statistical uncertainty $\sigma(r)$ after component separation. Doing so, we obtain the requirement shown in Eq. \ref{eq:req}: both central frequencies and relative gains should be calibrated, on all frequencies, to the $\sim1-2\%$ level. We have verified these results using two different component separation methods. We have also briefly studied the possibility of self-calibrating bandpass systematics by including them as parameters in the multi-frequency likelihood, informed by calibration priors. In this case, bandpass uncertainties produce a broadening of the posterior distribution, degrading the final constraints on most foreground parameters. For the foreground models explored here, however, the final constraints on $r$ remain relatively unaffected by the additional uncertainty, even assuming a $\sim5\%$ prior on the bandpass parameters. The success of this self-calibration method however depends crucially on the accuracy of the foreground model used in the multi-frequency likelihood, and therefore the target requirements for bandpass calibration should still be at the percent level.

Our findings are therefore encouraging in that current and future ground-based CMB experiments should be able to control bandpass-related systematics, and sufficiently mitigate their impact on the final constraints on primordial $B$-modes using techniques that are currently at our disposal.

Many other instrumental effects can contribute to the degradation of component separation methods and therefore limit our ability to reliably detect a non-zero tensor-to-scalar ratio.  These include the calibration of the polarization angles, beam mismatch, and instrumental polarization and cross-polarization. To achieve $r < 0.01$, these effects will undoubtedly have to be characterized with unprecedented precision. 

\section*{Acknowledgements}
\label{sec:acknowledgements}
We would like to thank Jo Dunkley from Princeton University, Johannes Hubmayr from NIST and Jeff McMahon from the University of Michigan for useful comments and discussions. David Alonso acknowledges support  from  the  Science  and  Technology  Facilities  Council (STFC) and the Beecroft Trust and Jonathan Ward acknowledges support from the NASA Space Technology Research Fellowship Program.

\bibliography{main}

\end{document}